\documentclass[%
 reprint,
 amsmath,amssymb,
 aps,
]{revtex4-2}

\usepackage{graphicx}% Include figure files
\usepackage{dcolumn}% Align table columns on decimal point
\usepackage{bm}% bold math
\usepackage{hyperref}% add hypertext capabilities
\hypersetup{colorlinks=true, allcolors={blue}}
\newcommand{\yn}{Ba$_3$YbB$_9$O$_{18}$}
\newcommand{\gt}{Ba$_3$GdB$_3$O$_9$}
\newcommand{\gn}{Ba$_3$GdB$_9$O$_{18}$}
\newcommand{\yt}{Ba$_3$YbB$_3$O$_9$}
\newcommand{\kbagd}{KBaGd(BO$_3$)$_2$}
\newcommand{\kbayb}{KBaYb(BO$_3$)$_2$}
\newcommand{\gdpo}{NaGdP$_2$O$_7$}
\newcommand{\ybpo}{NaYbP$_2$O$_7$}

\begin{document}

%\preprint{APS/123-QED}

\title{Sub 1\,K Adiabatic Demagnetization Refrigeration with Rare-Earth Borates Ba$_3$XB$_9$O$_{18}$ and Ba$_3$XB$_3$O$_9$, X = (Yb, Gd)}% Force line breaks with \\

\author{Marvin Klinger}
 \email{marvin.klinger@physik.uni-augsburg.de}
\author{Tim Treu}%

\author{Felix Kreisberger}

\author{Christian Heil}

\author{Anna Klinger}

\author{Anton Jesche}

\author{Philipp Gegenwart}

 \email{philipp.gegenwart@physik.uni-augsburg.de}
\affiliation{%
 Experimental Physics VI, Center for Electronic Correlations and Magnetism, Institute of Physics, University of Augsburg, 86159 Augsburg, Germany
}%


%\affiliation{
%\affiliation{
%\affiliation{%


\date{\today}% It is always \today, today,

\begin{abstract}
Adiabatic demagnetization refrigeration (ADR) is regaining relevance for the refrigeration to temperatures below 1 K as global helium-3 supply is increasingly strained. While ADR at these temperatures is long established with paramagnetic hydrated salts, more recently frustrated rare-earth oxides were found to offer higher entropy densities and practical advantages since they do not degrade under heating or evacuation.
We report structural, magnetic and thermodynamic properties of the rare-earth borates Ba$_3$XB$_9$O$_{18}$ and Ba$_3$XB$_3$O$_9$ with X = (Yb, Gd). Except for Ba$_3$GdB$_9$O$_{18}$, which orders at 108\,mK, the three other materials remain paramagnetic down to their lowest measured temperatures. ADR performance starting at 2\,K in a field of 5\,T is analyzed and compared to literature results.
\end{abstract}

\maketitle

%\tableofcontents

\section{Introduction}

Temperatures below one Kelvin are increasingly utilized not only in basic research but also in a growing number of industrial applications \cite{cao:2021}.
%Currently dilution refrigerators (DRs) are the generally preferred method of refrigeration for this temperature range.
%They offer high cooling power and continuous operation.
While dilution refrigerators have long been the default option for obtaining and maintaining low temperatures \cite{london:1962, pobell:2007a} there have been recent advances and renewed interest in adiabatic demagnetization refrigeration (ADR) \cite{zhao:2019, esat:2021, chen:2025, xu:2025}.
These can be grouped into two broad categories: refrigerator technology and refrigerant technology.

In terms of refrigerator technology there have been developments in continuous ADR which can maintain a set temperature by cycling multiple refrigerators connected via heat switches.
While such setups are more complex than single shot ADR, they offer the crucial benefit of maintaining continuous operation of the cold payload \cite{shirron:2010, shirron:2014, shirron:2018}. At the moment continuous ADR is not yet competitive with DRs in terms of cooling power \cite{duval:2020a}.

The field of refrigerant technology has long been limited by the performance characteristics of paramagnetic hydrated salts. Here, a large distance between magnetic moments, achieved by separating them with water molecules realizes very weak magnetic interaction and respectively low ordering temperatures. As a consequence, the magnetic entropy density and the cooling power are low. Furthermore, due to their chemical instability, paramagnetic hydrated salts like ferric ammonium alum, cerium magnesium nitrate and others need to be encapsulated for use in a vacuum environment and even then they may never be heated above room temperature. Otherwise the crystal water contained in these substances will quickly escape and the structural collapse makes them inoperative for ADR.

There have been three mayor development avenues for new refrigerants with higher entropy density and better chemical stability:
Single crystal refrigerants like gadolinium gallium garnet or ytterbium gallium garnet offer high entropy density and allow ADR to 1~and 0.2~K, respectively  \cite{paixaobrasiliano:2020, kleinhans:2023b, he:2025}.
Intermetallic compounds like YbNi$_{1.6}$Sn have a much higher thermal conductivity because of the conduction electron contribution, while maintaining a high entropy density and are thus a suitable alternative at temperatures down to $0.2\,\mathrm{K}$ \cite{tokiwa:2016, gruner:2024b, shimura:2025, zhang:2025}.
The third alternative are rare-earth oxides in which geometrical frustration reduces the ordering temperature, allowing much lower ADR end temperatures in combination with a high entropy density \cite{tokiwa:2021, jesche:2023, arjun:2023, arjun:2023c, koskelo:2023, lin:2024, telang:2025, treu:2025, berardi:2025, channarayappa:2025, cui:2025, lin:2025, wang:2025, xu:2025, yao:2025}.
These materials are chemically stable, the polycrystal synthesis can easily be upscaled and excellent thermal and mechanical properties are obtained by compressing them with fine silver powder admixture, while maintaining a relatively high entropy density. Prototypes of this new class of oxide ADR materials are the borates KBaX(BO$_3$)$_2$ with magnetic triangular lattices of X = Yb, Gd~\cite{sanders:2017, pan:2021}.
While \kbagd{} orders antiferromagnetically at 263\,mK, the ADR experiment on a pellet in the PPMS, starting at 2\,K in a field of 5\,T reveals a minimal temperature of 122\,mK~\cite{jesche:2023}. Assuming a scaling of the ordering temperatures by the square of the ratio of the saturation moments (which is about 30), for isostructural \kbayb{} a very low ordering near 9\,mK has been deduced and indeed the system remains paramagnetic in ADR experiments down to 16\,mK~\cite{tokiwa:2021}. This is much lower than the minimal temperatures that can be obtained with metallic ADR materials mentioned above. Both geometrical frustration due to the triangular magnetic lattice and K$^+$/Ba$^{2+}$ site randomness~\cite{sanders:2017} may be important with respect to the outstanding ADR performance of the two materials~\cite{jesche:2023}. 

\yt{} is another triangular lattice ADR candidate material with a very similar intralayer Yb-Yb distance (5.43 \AA) as in \kbayb{}, but an about 30\% larger interlayer Yb-Yb distance \cite{gao:2018, zeng:2020, bag:2021, cho:2021}, making it more two-dimensional. Furthermore, \yt{} has no site randomness, but does exhibit two inequivalent Yb sites. We also study triangular lattice \yn{}, with a significantly 
%A comparison between the two compounds (also for their Gd-versions with classical spin $S=7/2$) can reveal information on the impact of three-dimensional couplings to magnetism and ADR in this family of triangular-lattice magnets.
larger in-plane Yb-Yb distance of 7.18 \AA, which arises from a larger separation of YbO$_6$ octahedra by the triangular arrangement of the BO$_3$ groups in the structure \cite{cho:2021, khatua:2022, liu:2025}. It is interesting to compare the ADR performance of these materials, together with their Gd counterparts, to KBaX(BO$_3$)$_2$ (X=Yb, Gd).
In the Gd variants with classical spin $S=7/2$ the higher entropy density improves ADR performance at the expense of minimal ADR temperature, due to the enhanced magnetic couplings.

The paper is organized as follows. After the description of the utilized methods, we report below structural, magnetic, thermodynamic and ADR properties of Ba$_3$XB$_9$O$_{18}$ and Ba$_3$XB$_3$O$_9$ with X = (Yb, Gd). No indication of magnetic ordering has been found in both Yb-compounds down to the lowest obtained ADR temperatures of below 40\,mK. In \gt{} a broad maximum is found at low temperatures. In \gn{}, a Schottky-type broad peak, as well as a sharp phase transition, indicative of long range magnetic order at lower temperatures, are observed in the heat capacity.

\section{Materials and Methods \label{sec:materials-methods}}

Polycrystals of Ba$_3$XB$_9$O$_{18}$ and Ba$_3$XB$_3$O$_9$ with X = (Yb, Gd) were prepared by solid state reaction. All reagents were preheated, weighed and thoroughly ground in an agate mortar.
Subsequently they were preheated in a furnace under laboratory atmosphere at 700°C over 12\,h.
After the first furnace run the reagents were allowed to cool and reground.
A second oven run was performed again under laboratory atmosphere but this time with an increased temperature of 900°C over 24\,h.

The products were ground again and analyzed for phase purity by powder X-ray diffraction in an PANAlytical Empyrean XRD.
To ensure good thermal contact, even at very low temperatures, the substances were mixed with 50\% silver powder by weight and pressed into pellets for further analysis.

Pellets of 3\,mm diameter were prepared for specific heat and magnetization measurements while 15\,mm diameter pellets were prepared for direct magnetocaloric studies.

Magnetization was measured in a Quantum Design MPMS3 SQUID magnetometer; for measurements in the range $0.4\,\mathrm{K} < T < 2\,\mathrm{K}$ the $^3$He option was utilized.
The samples magnetic moment was measured under an isothermal field sweep with a vibrating sample magnetometer (VSM).
In order to achieve higher accuracy, DC magnetization measurements were conducted at the initial and final fields and the VSM data was scaled to these DC measurements.
Furthermore the data was corrected for the calculated geometry factor of the cylindrical samples.

Specific heat was measured in a Physical Property Measurement System (PPMS) DynaCool by Quantum Design. For measurements at lower temperatures, a $^3$He upgrade was utilized, for which a small piece was cut from the middle of a 3\,mm pellet in order to limit the total specific heat and thus allow for more accurate and faster measurements.

Actual ADR performance of the samples was determined in a custom built adiabatic demagnetization refrigeration setup inside a PPMS similar to the one described in\,\cite{tokiwa:2021, jesche:2023, arjun:2023c, treu:2025}.
This setup comprises a standard PPMS-puck that mounts a polyimide (PI) frame in which the pellet is supported by aromatic polyamide yarns.
The suspended Pellet is shielded from infrared radiation by a polished brass cap.
Thermometry in this setup is performed by an ultra low specific heat, low time constant custom chip thermometer made from a commercially available RuO$_2$ chip resistor by removing the contacts and substrate.
This thermometer was calibrated against a reference thermometer and attached to the pellet with GE 7031 varnish.
Care was taken as not to overheat the thermometer. Thus, the excitation was limited to less than $50\,\mathrm{fW}$.
Similarly a heater was prepared from a lower resistance chip and also attached to the pellet.
To limit thermal flux into the sample, superconducting wiring is utilized.

\section{Results}

\subsection{Structure}

Powder X-ray diffraction (PXRD) was performed at room temperature with a PANAlytical Empyrean diffractometer.
The diffraction pattern were compared to the ICCD PDF-5+ database \cite{kabekkodu:2024, ilyukhin:1993, khamaganova:1999} to determine phase purity.
All samples contain only small fractions of foreign phases of which the most common were identified as the corresponding rare earth borates $(\mathrm{Yb}, \mathrm{Gd})\mathrm{BO}_3$.
A central peak at $44.5\,$° can be attributed to the sample holder.

\begin{table*}%[b]%The best place to locate the table environment is directly after its first reference in text
\caption{Structural data for the compounds, verified by powder XRD.\label{tab:structure}}
\begin{ruledtabular}
\begin{tabular}{ccccc}
\textbf{Compound}	& \textbf{Lattice}	& \textbf{Space group} & \textbf{Unit cell parameters} & \textbf{Molar volume}\\
\colrule
\yn{} \cite{kabekkodu:2024}    & hexagonal    & $P6_3/m$ & $\text{a: }7.1740\,\mathring{\mathrm{A}}, \text{c: }16.915\,\mathring{\mathrm{A}}$  & $227.0\,\mathrm{cm^3mol^{-1}}$\\
\yt{} \cite{ilyukhin:1993} 		& hexagonal    & $P6_3cm$ & $\text{a: }9.3830\,\mathring{\mathrm{A}}, \text{c: }17.441\,\mathring{\mathrm{A}}$  & $133.5\,\mathrm{cm^3mol^{-1}}$\\
\gn{} \cite{kabekkodu:2024}	& hexagonal    & $P6_3/m$ & $\text{a: }7.1934\,\mathring{\mathrm{A}}, \text{c: }17.206\,\mathring{\mathrm{A}}$  & $232.2\,\mathrm{cm^3mol^{-1}}$\\
\gt{} \cite{khamaganova:1999}		& trigonal     & $R\bar{3}$ & $\text{a: }12.776\,\mathring{\mathrm{A}}, \text{c: }9.6063\,\mathring{\mathrm{A}}$& $141.8\,\mathrm{cm^3mol^{-1}}$\\
\end{tabular}
\end{ruledtabular}
\end{table*}

\begin{figure}[ht]
\includegraphics[width=\columnwidth]{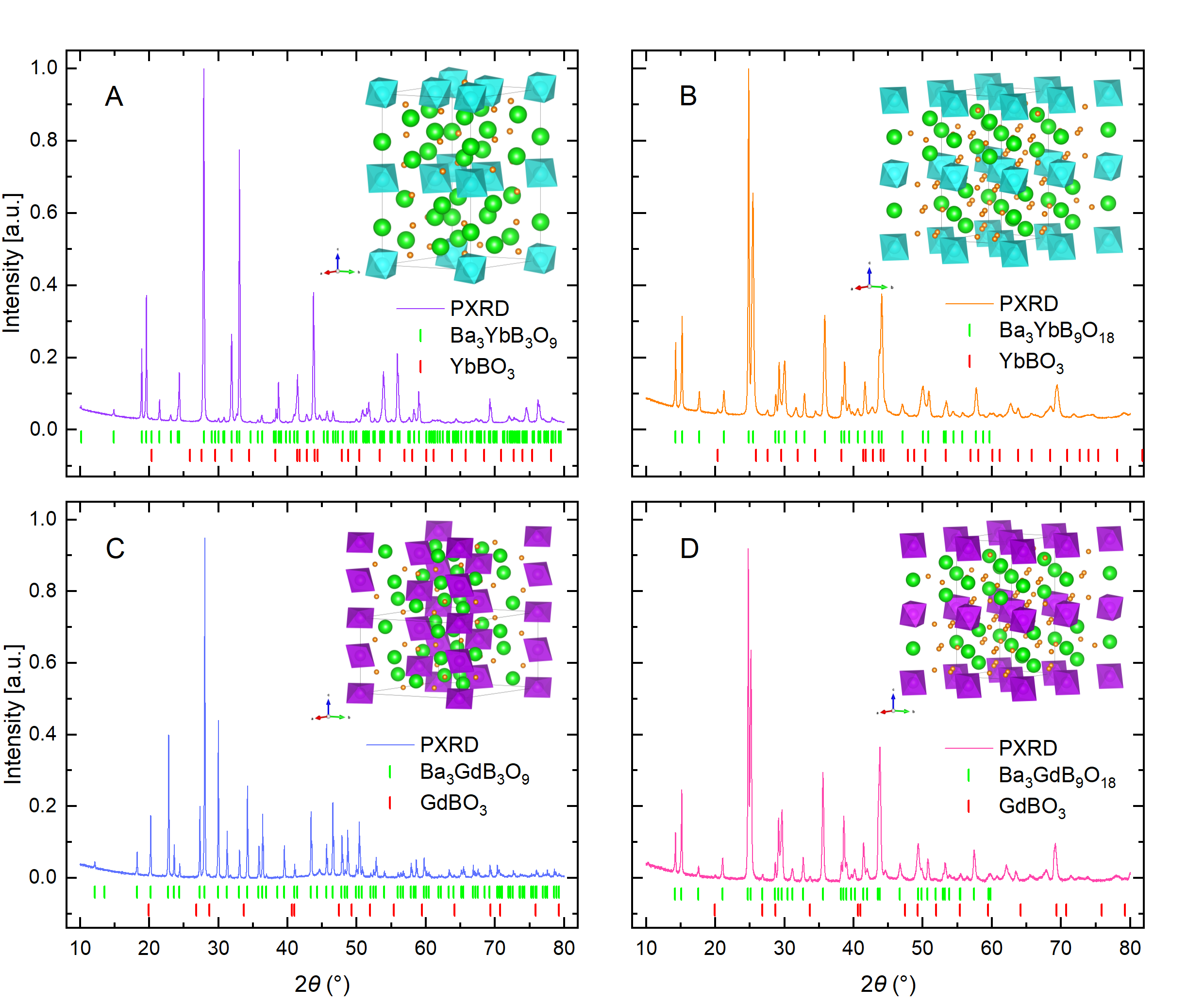}
\caption{PXRD data and literature peak positions for the product (green) as well as the most common foreign phase (red). The insets show the crystal structures with rare earth ions surrounded by oxygen octahedra (Yb: blue, Gd: purple) and separated by boron (orange) and barium (green). Oxygen atoms have been omitted for clarity and solid lines represent unit cells.\label{fig:pxrd}}
\end{figure}

%For these samples, all rare earth ions are situated in oxygen octahedra leading to CEF splitting of the electronic energy levels.
Both \yn{} and \gn{} form a hexagonal lattice of identical space group $P6_3/m$ with one rare earth site forming layers in the $ab$ plane (fig. \ref{fig:pxrd} B,D).
The rare earth ions are coordinated in a hexagonal fashion, allowing for geometric frustration \cite{gao:2018, zeng:2020}.
The layers differ by alternating orientations of the oxygen octahedra, further enhancing inter-layer frustrations.
Similar structure has been observed for Ba$_3$TbB$_3$O$_9$ \cite{kelly:2025}.

\yt{} forms a hexagonal lattice of space group $P6_3cm$ \cite{cho:2021}.
In this compound the ytterbium ions also form layers in the $ab$ plane, however these layers contain two alternating Yb sites and are distorted (fig. \ref{fig:pxrd} A).
The two sites share the orientation of the oxygen octahedra and the layers differ by alternating orientations of the octahedra along the $c$ axis.
This arrangement allows for additional frustration between the two distinct Yb sites.
\gt{} exhibits mayor differences to the other compounds.
It crystallizes in a trigonal lattice ($R\bar{3})$ with two distinct gadolinium sites forming alternating chains along the $c$ axis (fig. \ref{fig:pxrd} C).
These chains are separated from each other with nearest neighbor chains offset~along~$c$.
\mbox{Similar structures have been known} to harbor a manifold of magnetic phases in other substances \cite{bera:2014, zhou:2025}.
There is little literature available on \gt{} except for the structure, which we confirmed by PXRD \cite{kabekkodu:2024}.

% YbB3: two Yb sites minor difference in positon, layers of Yb in c, mayor difference in Oxygen octahedron between layers
% YbB9: one Yb site, layers of Yb in c, mayor difference in Oxygen octahedron between layers
% GdB3: two Gd sites, chains of alternating Gd in c, mayor difference in Oxygen octahedron along the chains
% GdB9: one Gd site, layers of Gd in c, mayor difference in Oxygen octahedron between layers
\begin{table}[ht]%The best place to locate the table environment is directly after its first reference in text
\caption{Distance between magnetic moments, verified by powder XRD.\label{tab:distance}}
\begin{ruledtabular}
\begin{tabular}{ccc}
\textbf{Compound}	& \textbf{Nearest neighbor}	& \textbf{Next n. neighbor}\\
\colrule
\yn{} \cite{kabekkodu:2024}    & $7.174\,\mathring{\mathrm{A}}$    & $8.458\,\mathring{\mathrm{A}}$\\
\yt{} \cite{ilyukhin:1993}     & $5.417\,\mathring{\mathrm{A}}$    & $8.721\,\mathring{\mathrm{A}}$\\
\gn{} \cite{kabekkodu:2024}	& $7.193\,\mathring{\mathrm{A}}$    & $8.603\,\mathring{\mathrm{A}}$\\
\gt{} \cite{khamaganova:1999}	& $4.776\,\mathring{\mathrm{A}}$    & $7.710\,\mathring{\mathrm{A}}$\\
\end{tabular}
\end{ruledtabular}
\end{table}

\subsection{Magnetic properties}

\begin{figure}[ht]
\includegraphics[width=\columnwidth]{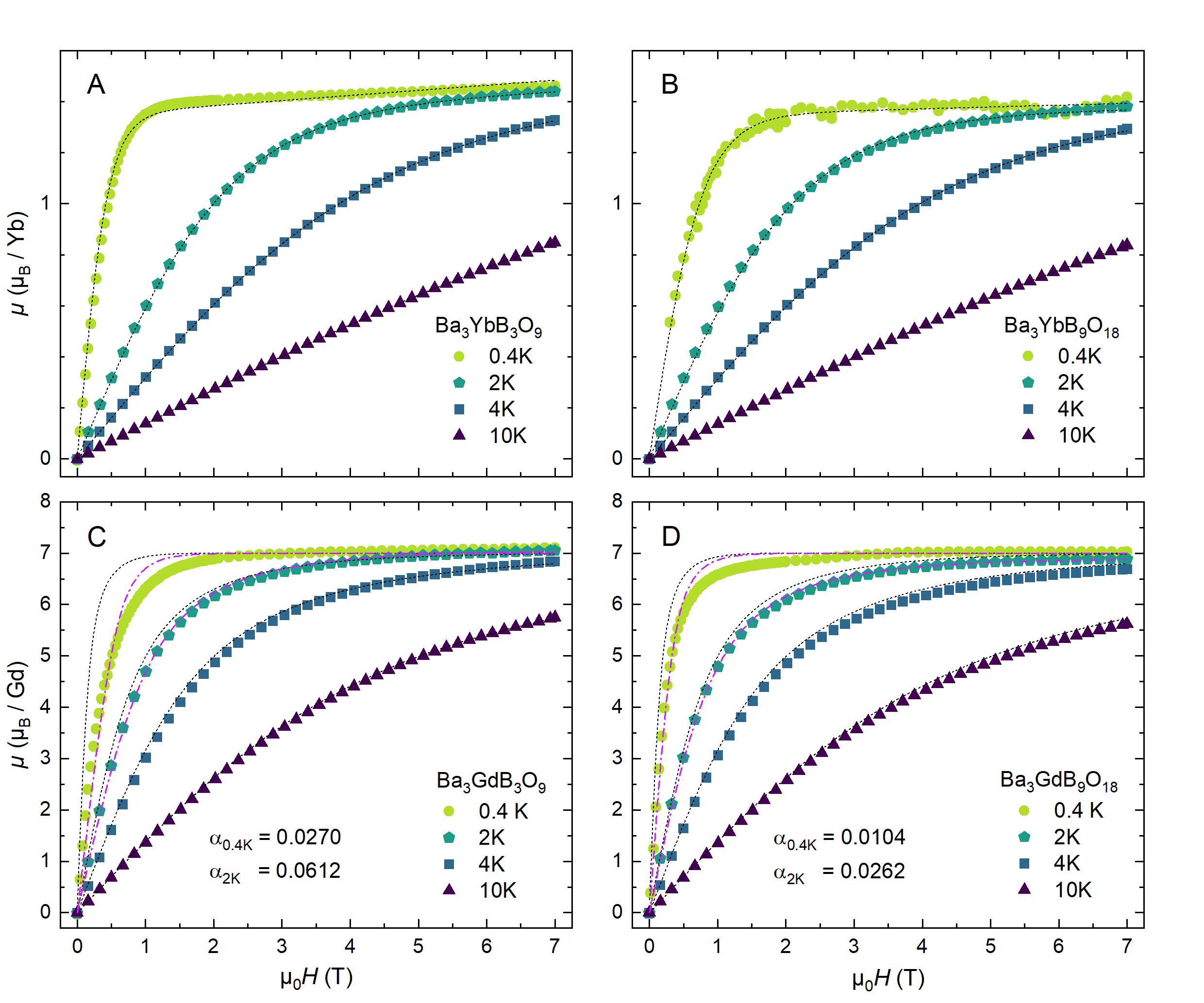}
\caption{Magnetic moment $\mu$ in units of Bohr magneton $\mu_\mathrm{B}$ per magnetic ion versus applied field $H$ for different temperatures for Ba$_3$YbB$_3$O$_9$ (A), Ba$_3$YbB$_9$O$_{18}$ (B),  Ba$_3$GdB$_3$O$_9$ (C) and Ba$_3$GdB$_9$O$_{18}$ (D). The black dotted lines show calculated moments from eq. \ref{eq:freeIon} for a free ion. The dashed pink lines show a mean field approximation according to eq. \ref{eq:meanField}. \label{fig:mh}}
\end{figure}

%Isothermal magnetization
Fig. \ref{fig:mh} displays our data of the isothermal magnetization of the four different materials at temperatures between 0.4 and 10 K.
The magnetic moment of non-interacting dipoles can be described by the Brillouin function:
%\begin{widetext}
%\begin{equation} \label{eq:freeIon}
%    \mu(B) = gJ \left[\frac{2J + 1}{2J}\coth\left(\frac{2J + 1}{2J} \frac{g \mu_\mathrm{B}  J  B}{k_\mathrm B T} \right) - \frac{1}{2J}\coth\left(\frac{g \mu_\mathrm{B} B}{2k_\mathrm B  T}\right)\right] + x_0B
%\end{equation}
%\end{widetext}
\begin{eqnarray} \label{eq:freeIon}
\mu(B) =&& gJ \left[\frac{2J + 1}{2J}\coth\left(\frac{2J + 1}{2J} \frac{g \mu_\mathrm{B}  J  B}{k_\mathrm B T} \right)\right.\nonumber\\
&& \left.- \frac{1}{2J}\coth\left(\frac{g \mu_\mathrm{B} B}{2k_\mathrm B  T}\right)\right] + x_0B
\end{eqnarray}
with a total angular momentum of $J$ and the Landé~factor $g$.
The function $\mu(B)$ (eq. \ref{eq:freeIon}) was fitted to the data with fixed $T$ and fixed $J = 7/2$ and $g = 2$ for the Gd compounds.
A van Vleck contribution $x_0B$ was also added although this does not become dominant for any of the materials.

The ytterbium samples (fig. \ref{fig:mh} panels A, B), can be approximated as free-ion systems, as the magnetic interaction energies are very small compared to thermal energy at the studied temperatures; note that their magnetic interaction is much smaller compared to their Gd counterparts due to the smaller effective moment.
For these samples saturation was achieved in the $0.4\,\mathrm{K}$ and $2\,\mathrm{K}$ measurements.
Fitting eq. \ref{eq:freeIon} yields a $g$ factor of $2.71$ for \yt{} and $2.70$ for \yn{}.
Significant anisotropy of $g$ has been observed previously for similar Yb compounds \cite{ma:2018} and for \yt{} by Bag~et~al.~\cite{bag:2021} ($g_{ab}~=~2.19, g_c=3.38$) but the polycrystalline average observed is in good agreement with reports by Cho~et~al.~\cite{cho:2021} ($g_\mathrm{PM} = 2.77(1)$).

Both gadolinium systems saturate close to the expected $\mu_\mathrm{sat} = 7\mu_\mathrm{B}$ for high fields and low temperatures indicating the expected $g=2$.
Note, that the single-ion approximation (black, fig.\ref{fig:mh} C, D) is inaccurate due to the interaction between ions for the data below 2~K. This manifests as a shift of the magnetic saturation to higher fields than expected in the free ion model.
This effect is enhanced in \gt{} due to the higher magnetic moment density.
Although the structural parameters are similar, the interaction in the gadolinium compounds is enhanced when compared to the ytterbium compounds, as the magnetic moments are significantly larger ($\mathrm{S_{eff}}=1/2$ vs. $\mathrm{S}=7/2$).
An increase in interactions upon rare earth replacement has been observed in the iso structural \kbayb{} \cite{tokiwa:2021} and \kbagd{} \cite{jesche:2023} as well as in the iso structural \ybpo{} \cite{arjun:2023c} and \gdpo{} \cite{telang:2025}.
Therefore, a mean field approximation was conducted by fitting eq. \ref{eq:meanField} to the $2\,\mathrm{K}$ and $0.4\,\mathrm{K}$ gadolinium datasets (pink, fig. \ref{fig:mh} C, D):
\begin{equation} \label{eq:meanField}
    \mu = \mu_\mathrm{sat}B_J\left[ \frac{g\mu_\mathrm{B}J}{k_\mathrm{B}T}\left(H+\alpha \mu \right) \right].
\end{equation}
The mean field model fits the data significantly better than the free ion approximation, however there are still significant deviations, likely due to CEF effects (see below).
For $2\,\mathrm{K}$ the mean field model fits significantly better than for $0.4\,\mathrm{K}$.
The mean field parameter $\alpha$ appears to be temperature dependent as both samples exhibit a decrease in $\alpha$ for lower temperatures.
%This is unexpected but could be related to the onset of magnetic order at temperatures below $0.4\,\mathrm{K}$ or the effect of CEF splitting.
This unexpected temperature dependence hints at influences that can not be described well by the mean field model - likely CEF effects.

\begin{figure}[ht]
\includegraphics[width=\columnwidth]{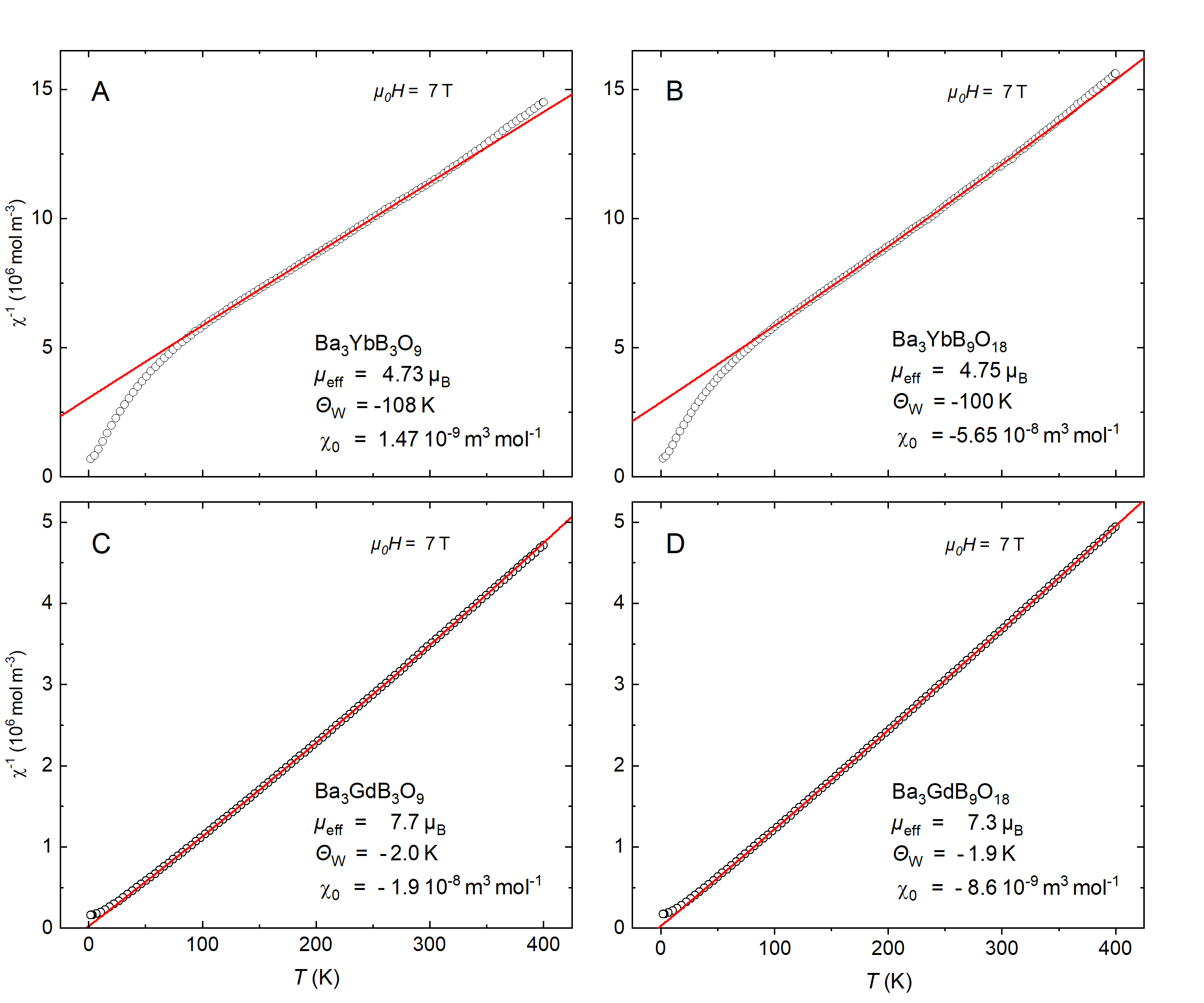}
\caption{Inverse magnetic susceptibility at an applied field of $7\,\mathrm{T}$ and fitted with eq. \ref{eq:invChi} with the listed parameters for Ba$_3$YbB$_3$O$_9$ (A), Ba$_3$YbB$_9$O$_{18}$ (B),  Ba$_3$GdB$_3$O$_9$ (C) and Ba$_3$GdB$_9$O$_{18}$ (D). \label{fig:mt}}
\end{figure}

% Susceptibility
The magnetic susceptibility $\chi$ of the samples was measured down to $0.4\,\mathrm{K}$.
For high temperatures, an external field of $7\,\mathrm{T}$ was applied and for low temperatures, a low field of $5\,\mathrm{mT}$ was utilized.
The inverse susceptibility was fitted with a modified Curie~Weiss function (eq. \ref{eq:invChi}) to obtain effective moments and Weiss temperature \cite[eq. 1.18]{balanda:2017}:
\begin{equation} 
    \chi^{-1} = (T-\Theta_\mathrm{W})\left[ 3k_\mathrm{B}\mu_0N_\mathrm{A}\mu_\mathrm{B}^2 \mu_\mathrm{eff}^{-2} + \chi_0\right] \label{eq:invChi}.
\end{equation}
For high temperatures, Yb and Gd samples show a linear relation between $\chi^{-1}$ and $T$.
For $T<100\,\mathrm{K}$ both Yb compounds deviate from the linear relation, as the CEF splitted energy levels occupation shifts \cite{sala:2019}.
At low temperatures and low magnetic field, the Yb samples exhibit Kramers behavior thus returning to a linear $\chi^{-1}(T)$.
To obtain $\mu_\mathrm{eff}$ and $\Theta_\mathrm{W}$ for the Yb samples eq. \ref{eq:invChi} was fitted between $100\,\mathrm{K}$ and $300\,\mathrm{K}$ in fig. \ref{fig:mt}.
For \yt{} the resulting effective moment of $4.73\,\mu_\mathrm{B}$ and Weiss temperature of $\Theta_\mathrm{W}=-108\,\mathrm{K}$ are close to reports by Gao~et~al.~\cite{gao:2018} ($\mu_\mathrm{eff}=4.60\,\mu_\mathrm{B}, \Theta_\mathrm{W} = -109\,\mathrm{K}$) as well as by Cho~et~al.~\cite{cho:2021} ($\mu_\mathrm{eff} = 4.89(1)\,\mu_\mathrm{B}, \Theta_\mathrm{W}=-105(1)\,\mathrm{K}$).
For \yn{} the effective moment of $4.75\,\mu_\mathrm{B}$ and Weiss temperature $\Theta_\mathrm{W}=-100\,\mathrm{K}$ are close to data by Khatua~et~al.~\cite{khatua:2022} ($\mu_\mathrm{eff}=4.73\,\mu_\mathrm{B}, \Theta_\mathrm{W} = -90\,\mathrm{K}$) as well as by Cho~et~al.~\cite{cho:2021} ($\mu_\mathrm{eff} = 4.74(1)\,\mu_\mathrm{B}, \Theta_\mathrm{W} = -99(1)\,\mathrm{K}$). There is a small discrepancy to the observations by Liu~et~al.~\cite{liu:2025} ($\mu_\mathrm{eff} = 4.34\,\mu_\mathrm{B}, \Theta_\mathrm{W} = -78.09\,\mathrm{K}$) but this dataset was collected at $0.2\,\mathrm{T}$ while our data was collected at a significantly higher field of $7\,\mathrm{T}$.
The resulting effective moments of $4.73\,\mu_\mathrm{B}$ and $4.75\,\mu_\mathrm{B}$ are close to the theoretical $\mu_\mathrm{eff} = g_J\sqrt{J(J+1)}\mu_\mathrm{B} = 4.54\,\mu_\mathrm{B}$ for a free Yb$^{3+}$ ion with $J = 7/2$ and $g_J = 8/7$.
The absolute values of the Weiss temperatures from the high-temperature fit of the two Yb materials cannot be interpreted as indicator of large coupling between the moments, because they result from the crystal field splitting, leaving only the low lying doublet at low temperatures \cite{gao:2018}.
For \yn{} the energy gap of the effective two level system has been reported at $\Delta E / k_\mathrm{B} = 210.61\,\mathrm{K}$ \cite{liu:2025}.
%These results are in agreement with literature for \yt{} \cite{gao:2018, bag:2021} and \yn{} \cite{cho:2021, liu:2025} as well as similar Yb$^{3+}$ based oxides \cite{sanders:2017}.

The Gd samples show a linear relation for $\chi^{-1}(T)$ down to approximately $25\,\mathrm{K}$ when polarization of the moments starts to become significant - therefore fitting was performed between $100\,\mathrm{K}$ and $350\,\mathrm{K}$.
The resulting effective moments are close to the theoretical $\mu_\mathrm{eff}~=~7.94\,\mu_\mathrm{B}$.
For \gt{} the effective moment of $7.7\,\mu_\mathrm{B}$ and Weiss temperature of $\Theta_\mathrm{W}=-2.0\,\mathrm{K}$ are comparable to reports on the similar frustrated oxide magnet \kbagd{} by Sanders~et~al.~\cite{sanders:2017} ($\mu_\mathrm{eff} = 7.70\,\mu_\mathrm{B}, \Theta_\mathrm{W} = -1.64\,\mathrm{K}$) as well as by Jesche~et~al.~\cite{jesche:2023} ($\mu_\mathrm{eff} = 7.91\,\mu_\mathrm{B}, \Theta_\mathrm{W} = -1.4\,\mathrm{K}$).
Notably, we observe a slightly negative $\Theta_\mathrm{W}$ even in the high temperature fit.
For \gn{} the resulting effective moment of $7.3\,\mu_\mathrm{B}$ is slightly lower than the theoretical value for a free ion model. Even though nearest neighbor distances are quite large in this compound the reducton in $\mu_\mathrm{eff}$ could hint at interactions between the moments.
The Weiss temperature of $\Theta_\mathrm{W}=-1.9\,\mathrm{K}$ is quite close to zero, indicating a weak antiferromagnetic exchange.
These results also are in good agreement with other Gd$^{3+}$ frustrated oxides \cite{telang:2025}.
All compounds exhibit negative $\Theta_\mathrm{W}$ in the high temperature fits suggesting dominant antiferromagnetic interaction.

\begin{figure}[ht]
\includegraphics[width=\columnwidth]{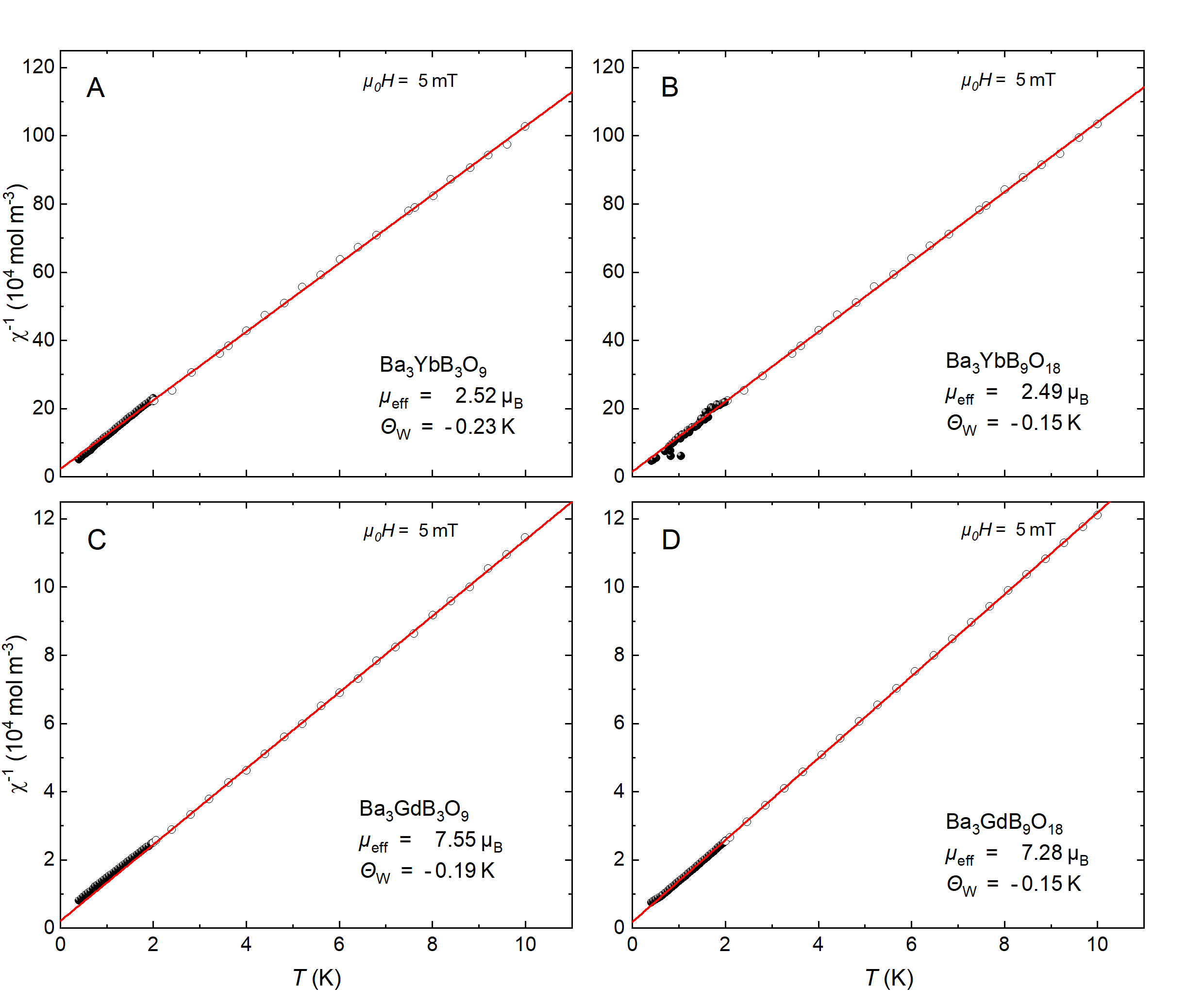}
\caption{Inverse susceptibility for low temperatures was measured at an applied field of $5\,\mathrm{mT}$, empty symbols were measured with helium-4, filled symbols with helium-3. This data was fitted with eq.~\ref{eq:invChi} with $\chi_0 = 0$.\label{fig:invChi_lt}}
\end{figure}

For low temperatures and low fields, the $\chi^{-1}(T)$ behavior of all samples is linear.
The Yb compounds now indicate fluctuating $\mathrm{S} = 1/2$ Kramers doublets with reduced $\mu_\mathrm{eff}$ as compared to high temperatures.

For \yt{} the effective moment of $2.52\,\mu_\mathrm{B}$ is very close to reports by Gao~et~al.~\cite{gao:2018} ($\mu_\mathrm{eff} = 2.61\,\mu_\mathrm{B}, \Theta_\mathrm{W} = -2.3\,\mathrm{K}$) as well as by Cho~et~al.~\cite{cho:2021} ($\mu_\mathrm{eff} = 2.56(1)\,\mu_\mathrm{B}, \Theta_\mathrm{W} = -0.24(1)\,\mathrm{K}$).
The obtained Weiss temperature $\Theta_\mathrm{W}=-0.23\,\mathrm{K}$ is close to the results by Cho~et~al.~\cite{cho:2021}, indicating very weak or highly frustrated interactions.

For \yn{} the effective moment of $2.49\,\mu_\mathrm{B}$ and Weiss temperature $\Theta_\mathrm{W}=-0.15\,\mathrm{K}$ are very close to data by Khatua~et~al.~\cite{khatua:2022} ($\mu_\mathrm{eff} = 2.32\,\mu_\mathrm{B}, \Theta_\mathrm{W} = -0.12\,\mathrm{K}$). Reports by Cho~et~al.~\cite{cho:2021} ($\mu_\mathrm{eff} = 2.31(1)\,\mu_\mathrm{B}, \Theta_\mathrm{W} = -0.077(2)\,\mathrm{K}$) suggest a lower $\Theta_\mathrm{W}$ but are based on measurements to $2\,\mathrm{K}$.

Both Gd samples continue to exhibit a high effective magnetic moment of $7.55\,\mu_\mathrm{B}$ and $7.28\,\mu_\mathrm{B}$ at low temperatures.
This is expected as the full $S=7/2$ multiplet contributes to the magnetic moment in Gd$^{3+}$ ions \cite{sanders:2017, jesche:2023, telang:2025}.

$\Theta_\mathrm{W}$ for all samples is increased in this temperature range, but remains negative.
This suggests antiferromagnetic behavior with low ordering temperature $T_\mathrm{N}$.
Both $\mathrm{Ba}_3\mathrm{X}\mathrm{B}_9\mathrm{O}_{18}$ samples show higher $\Theta_\mathrm{W}$ than their respective $\mathrm{Ba}_3\mathrm{X}\mathrm{B}_3\mathrm{O}_{9}$ counterparts.
The increase in $\Theta_\mathrm{W}$ corresponds with a lower magnetic moment density and thus reduced magnetic interactions.
All compounds show no signs of long-range order down to the lowest observed temperatures.

\subsection{Specific heat}

Specific heat was measured on the powder pressed samples with silver admixture in the PPMS.
The helium-3 data was scaled to the helium-4 data and the contributions of silver as well as the phonons were subtracted.
The phonon contribution was determined by fitting the high temperature data.
For low temperatures, the microcalorimeter measurements were extended below $600\,\mathrm{mK}$ by means of ADR warm up analysis.
In this process a pellet is allowed to warm up to the starting temperature $T_0$ after a magnetic cooldown under parasitic heat load.
From this warming curve $T(t)$ the specific heat $c_p$ can be calculated by:
\begin{equation}
c_p = \frac{m_\mathrm{mol}}{ m }\frac{d Q}{d T} = \frac{m_\mathrm{mol}}{m} \frac{dQ}{dt}\left( \frac{dT}{dt} \right)^{-1}.
\end{equation}
It has previously been shown that for the PPMS puck based setup the assumption of constant $\dot{Q}_\mathrm{parasitic}(T)$ holds for $20\,\mathrm{mK} < T < 1.5\,\mathrm{K}$ \cite{jesche:2023, telang:2025, wang:2025}.
The specific heat from the ADR measurements was scaled to the zero field data measured by microcalorimetry to determine the heat load.

\begin{figure}[ht]
\includegraphics[width=\columnwidth]{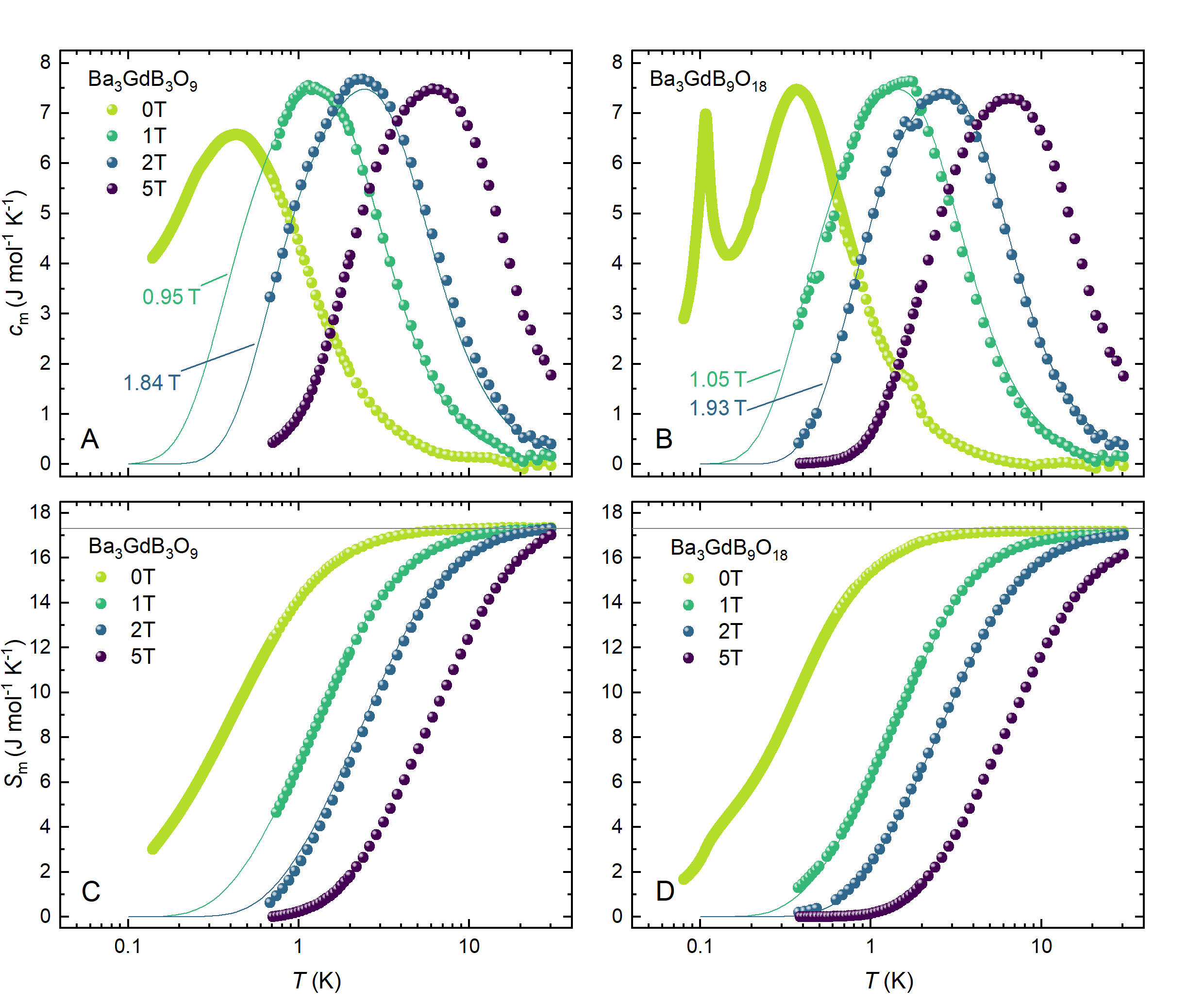}
\caption{Specific heat and magnetic entropy for \gt{}(A,C) and \gn{} (B,D) in various different fields. Colored lines display model calculations from eq.~\ref{eq:hcGd} with "effective" field values as indicated.
\label{fig:gdHc}}
\end{figure}

Magnetic specific heat $c_\mathrm{m}$ for a system of non interacting magnetic dipoles in an external field can be described as \cite[eq. 9.15b]{pobell:2007}:
\begin{eqnarray} \label{eq:hcGd}
 c_\mathrm{m} = && \frac{x^2R}{4}\sinh^{-2}\left( \frac{x}{2} \right)\nonumber\\
&& - \frac{x^2}{4}(2J+1)^2 \sinh^{-2}\left( \frac{x}{2}(2J+1) \right)\nonumber\\
\text{with} \quad x =&& \frac{\mu_\mathrm{B}gH_\mathrm{eff}}{k_\mathrm{B}T}.
\end{eqnarray}
The contribution of internal fields and antiferromagnetic exchange is considered through a effective field $H_\text{eff}$ at the position of each magnetic ion.
This effective field is determined through fitting the specific heat peak.
The values of $H_\mathrm{eff}$ are mostly lower than the applied field, which may be attributed to the antiferromagnetic exchange in the compounds, while a slightly enhanced effective field was used for \gn{} at an applied field of $1\,\mathrm{T}$.
%For a two level system with $S = J = 1/2$ this yields a Schottky type anomaly:
%\begin{equation}
%c_\mathrm{m} = x^2 R k_\mathrm{B}  \frac{ \exp{x} }{ \left( \exp{x}+1 \right)^2 } \qquad \text{with} \quad x = \frac{\mu_\mathrm{B}gH_\text{eff}}{k_\mathrm{B}T}. \label{eq:hcYb}
%\end{equation}
The specific heat data was fitted with eq. \ref{eq:hcGd}.
\gt{} shows a clear maximum in specific heat that is shifted to higher temperatures with increased field. In zero field also a broad peak is found with no indication of a long range order magnetic phase transition. 
%While the peak shape is similar for finite field the zero field specific heat curve significantly deviates from the non interacting model.
%This might be due to non negligible interaction between the $\mathrm{Gd}^{3+}$ ions.

In finite field the molar specific heat of \gn{} behaves  similar to that of \gt{}, as expected, also exhibiting a broad peak that is shifted to higher $T$ with increasing field. The zero field the specific heat shows a clear phase transition anomaly at $108\,\mathrm{mK}$, followed by a broader peak at $368\,\mathrm{mK}$.
%Interestingly, the FWHM of both peaks combined matches the expected value according to eq. \ref{eq:hcGd} as shown by the pink curve in fig. \ref{fig:gdHc} panel B.
%The broader peak matches a Shottky type anomaly with $S = 1/2$ and can be fitted by eq. \ref{eq:hcYb}.

%We propose that the broader peak is due to an effective two level system formed by the lowest occupied doublet of the Gd$^{3+}$ Ions. This doublet is split by the CEF leading to a Shottky type specific heat anomaly.
%Similar $S=1/2$ Schottky behavior has previously been reported for Gadolinium based hydrated salts $\mathrm{GdCl}_3\cdot6\mathrm{H}_2\mathrm{O}$ and $\mathrm{Gd}_2(\mathrm{SO}_4)_3\cdot8\mathrm{H}_2\mathrm{O}$ \cite{bogle:1962, wielinga:1967a}.
%The influence of CEF and structural distortions on the ground state splitting and single ion nature of Gd$^{3+}$ has been demonstrated for multiple compounds \cite{oyeka:2022, petersen:2023}.
The lambda-shaped peak indicates most likely antiferromagnetic order, given the negative sign of the Curie-Weiss temperature.
%at $T_\mathrm{N} = 108\,\mathrm{mK}$ indicates a long range magnetic ordering, which we determine to be antiferromagnetic due to the negative $\Theta_\mathrm{W}$.
The ordering occurs at a temperature close to, but lower than $|\Theta_\mathrm{W}| = 0.13\,\mathrm{K}$ which can be attributed to frustration in the magnetic lattice.

The origin of the absence of a magnetic transition in \gt{} is unclear, as this material crystallizes in a different structure with two inequivalent Gd sites and chain motives. It can also not be excluded, that an ordering would occur at temperatures lower than the minimal temperature of the ADR experiment.
%we do not observe the lambda-like ordering peak but rather a broadened single peak.
%This can be attributed to the increased frustration in the trigonal lattice as compared to the hexagonal lattice.

From the measured specific heat data magnetic entropy $S_\mathrm{m}$ was calculated using
\begin{equation}
    S_\mathrm{m} =  S_0 + \int dT \frac{c_\mathrm m}{T}.
\end{equation}
The offset constant $S_0$ is necessary, as $c_\mathrm{m}$ is only available starting from finite $T_\mathrm{min}$.
This offset was determined from magnetization data at $2\,\mathrm{K}$ by:
\begin{equation}
    S_\mathrm{m} = \mu_0\int_{B_0}^{B_1}dH\frac{\partial M}{\partial T}.
\end{equation}
The resulting magnetic entropy shows a clear saturation effect at the expected value of $R\ln{8}=17.3\,\mathrm{Jmol}^{-1}\mathrm{K}^{-1}$.
\gt{} saturates at slightly lower temperatures for a given field when compared to \gn{}.
The contribution to $S_\mathrm{m}$ of the lambda like peak in $c_\mathrm{m}$ for \gn{} is small when compared with the broader peak at higher temperatures. This clearly indicates that the ordered moment comprises only a low fraction of the total Gd moment.
%Compared to the Schottky behavior the breaking of long range order only contributes a small fraction of $S_\mathrm{m}$.

\subsection{Actual refrigeration performance}

\begin{figure}[ht]
\includegraphics[width=\columnwidth]{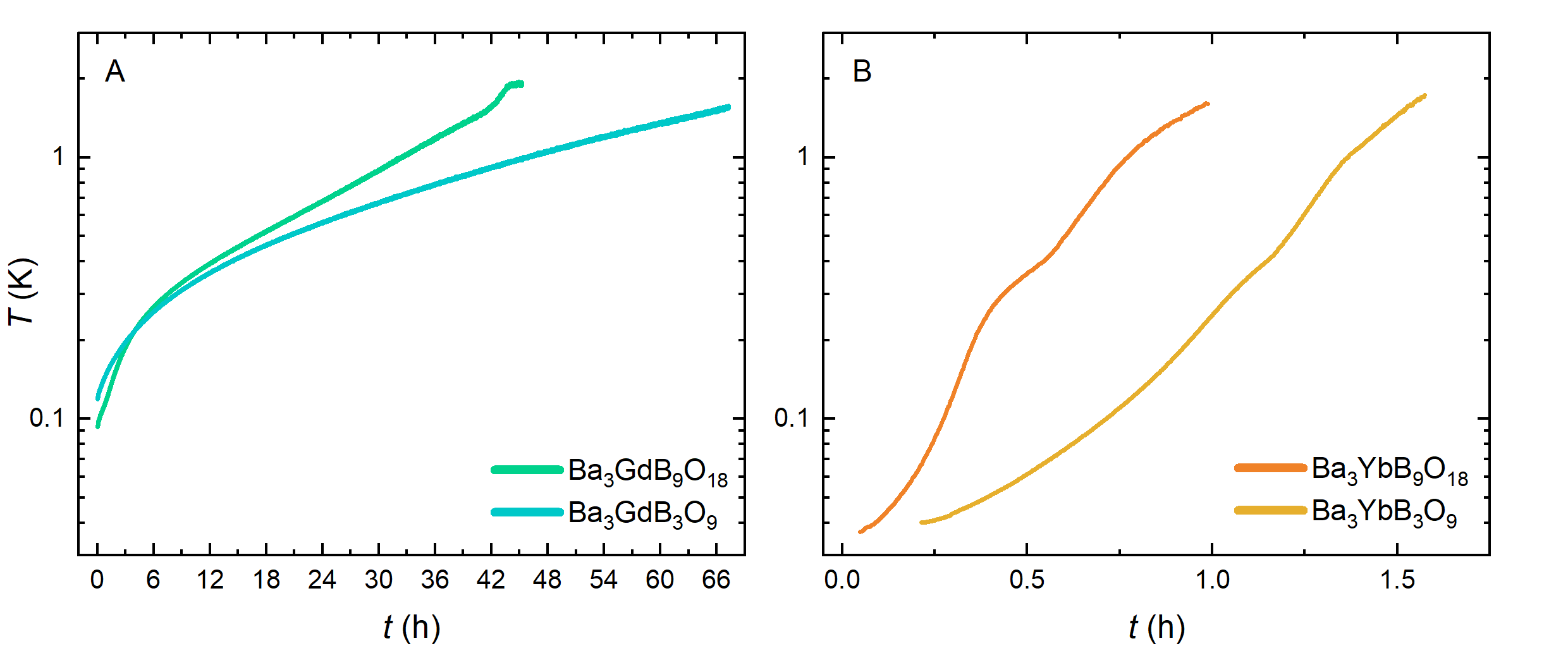}
\caption{Temperature of samples during parasitic warmup after ADR. Panels A and B show data for the Gd- and Yb-based compounds, respectively.\label{fig:adr_temperature}}
\end{figure}

To ascertain actual refrigeration performance of the compounds the $15\,\mathrm{mm}$ pellets were individually mounted to the PPMS-puck based setup described in chapter \ref{sec:materials-methods}.
%A calibrated, low thermal mass ruthenium oxide thermometer was attached to the pellet with GE/IMI-7031 varnish.
%Care was taken as not to overheat the thermometer, excitation was limited to less than $50\,\mathrm{fW}$.
This setup was isothermally magnetized to an applied field of $5\,\mathrm{T}$ at a starting temperature of $2\,\mathrm{K}$.
After magnetization the remaining helium atmosphere was evacuated from the sample chamber establishing near adiabatic conditions.
Subsequently the field was fully ramped down at a rate of $5\,\mathrm{mT/min}$.
The sample temperature was recorded during the parasitic warming back to the sample chamber temperature (fig. \ref{fig:adr_temperature}).
From the warmup curve $T(t)$ the refrigerant capacity $q$ is calculated by integrating the known parasitic heat flow.
It represents the amount of heat the refrigerant can accept when warming to a specified temperature:
\begin{equation}
    q(T) = \int_{t(T_{\mathrm{min}})}^{t(T)} \frac{dQ}{dt}dt' \label{eq:ref-cap}.
\end{equation}
Fig. \ref{fig:adr_temperature} displays the warmup data for all studied materials. The Yb compounds achieved the lowest ADR temperatures $T_\mathrm{ADR}$ with \yt{} attaining $40.0\,\mathrm{mK}$ and \yn{} attaining $37.0\,\mathrm{mK}$.
This is comparable to other frustrated oxides like \kbayb{} (Tokiwa~et~al. $40\,\mathrm{mK}$ \cite{tokiwa:2021}) or \ybpo{} (Arjun~et~al. $45\,\mathrm{mK}$ \cite{arjun:2023c}).
There is no indication for long range magnetic order in the Yb compounds down to the minimum temperatures attained during ADR, corroborating previous measurements \cite{khamaganova:1999,zeng:2020,bag:2021}. This suggests, that the obtained minimal temperatures are limited by the non-perfect adiabaticity of the setup and not yet by the vanishing entropy in zero field.
\yt{} warms 60\% slower than \yn{}, as expected, since the former has a 71\% higher magnetic moment density and 82\% higher magnetic entropy density compared to the latter.
This is also reflected in the 120\% enhanced refrigerant capacity~$q$.
%While the effective magnetic moment for both compounds is similar the effective magnetic moment density $\mu_\mathrm{eff}/V$ is significantly increased ($+72\%\,\mu_\mathrm{eff}/V$) for \yt{}.
\begin{table*}[t]%[b]%The best place to locate the table environment is directly after its first reference in text
\caption{Comparison of values for the effective magnetic moment and Curie-Weiss temperature (from fig.~4), minimal ADR temperature $T_\mathrm{ADR}$ for a full demagnetization from $2\,\mathrm{K}, 5\,\mathrm{T}$ and subsequent warm up back to $2\,\mathrm{K}$, refrigerant capacity $q$ for a warm up from $T_\mathrm{ADR}$ back to $2\,\mathrm{K}$, as well as full entropy density, calculated from the magnetic ion density, for the four different studied materials together with literature results for KBaYb(BO$_3$)$_2$ and KBaGd(BO$_3$)$_2$.
\label{tab:summary}}
\begin{ruledtabular}
\begin{tabular}{cccccc}
\textbf{Compound}	& $\mu_\mathrm{eff}$	& $\Theta_\mathrm{W}$ & $T_{\mathrm{ADR}}$ & $q$ & $\Delta S_\mathrm{m}$\\
\colrule
\yn{}		                            & $2.49\,\mu_\mathrm{B}$    &  $-0.15\,\mathrm{K}$ & $37.0\,\mathrm{mK}$    & $0.84\,\mathrm{mJcm}^{-3}$  & $24.0\,\mathrm{mJ}\mathrm{K}^{-1}\mathrm{cm}^{-3}$\\
\yt{}		                            & $2.52\,\mu_\mathrm{B}$    &  $-0.23\,\mathrm{K}$ & $40.0\,\mathrm{mK}$    & $1.78\,\mathrm{mJcm}^{-3}$  & $43.6\,\mathrm{mJ}\mathrm{K}^{-1}\mathrm{cm}^{-3}$ \\
\kbayb{} \cite{tokiwa:2021, treu:2025} & $2.28\,\mu_\mathrm{B}$    &  $-0.06\,\mathrm{K}$ & $40\,\mathrm{mK}$      & -                           & $57.95\,\mathrm{mJ}\mathrm{K}^{-1}\mathrm{cm}^{-3}$ \\
\gn{}		                            & $7.59\,\mu_\mathrm{B}$    &  $-0.13\,\mathrm{K}$ & $94.3\,\mathrm{mK}$    & $29.4\,\mathrm{mJcm}^{-3}$  & $66.8\,\mathrm{mJ}\mathrm{K}^{-1}\mathrm{cm}^{-3}$ \\
\gt{}		                            & $7.55\,\mu_\mathrm{B}$    &  $-0.19\,\mathrm{K}$ & $119.2\,\mathrm{mK}$   & $56.0\,\mathrm{mJcm}^{-3}$  & $104\,\mathrm{mJ}\mathrm{K}^{-1}\mathrm{cm}^{-3}$ \\
\kbagd{} \cite{jesche:2023, treu:2025} & $7.55\,\mu_\mathrm{B}$    &  $-0.55\,\mathrm{K}$ & $122\,\mathrm{mK}$     & -                           & $139\,\mathrm{mJ}\mathrm{K}^{-1}\mathrm{cm}^{-3}$ \\
\end{tabular}
\end{ruledtabular}
\end{table*}

As expected from the fact that the Gd materials, compared to their Yb counterparts feature a stronger magnetic interaction and thus reduction of entropy at higher temperatures, they exhibit higher $T_\mathrm{ADR}$ with \gt{} attaining $119\,\mathrm{mK}$ and~\gn{} \mbox{attaining~$94.3\,\mathrm{mK}$.}
%The ratio of $T_\mathrm{ADR}$ is increased for the Gd samples ($-20\%$\,$T_\mathrm{ADR}$) when compared to the Yb samples ($-8\%$\,$T_\mathrm{ADR}$).
While the difference in minimal temperature $T_\mathrm{ADR}$ is quite small for the two Yb compounds at $\Delta T_\mathrm{ADR, Yb} = 3.0\,\mathrm{mK}$ the difference in minimal temperature is increased between the Gd materials at $\Delta T_\mathrm{ADR, Gd} = 24.7\,\mathrm{mK}$.
\gt{} warms 60\% slower to 2\,K than \gn{} compatible with the 56\% higher entropy density.
This is also reflected in 105\% higher refrigerant capacity $q$.
\begin{figure}
\includegraphics[width=\columnwidth]{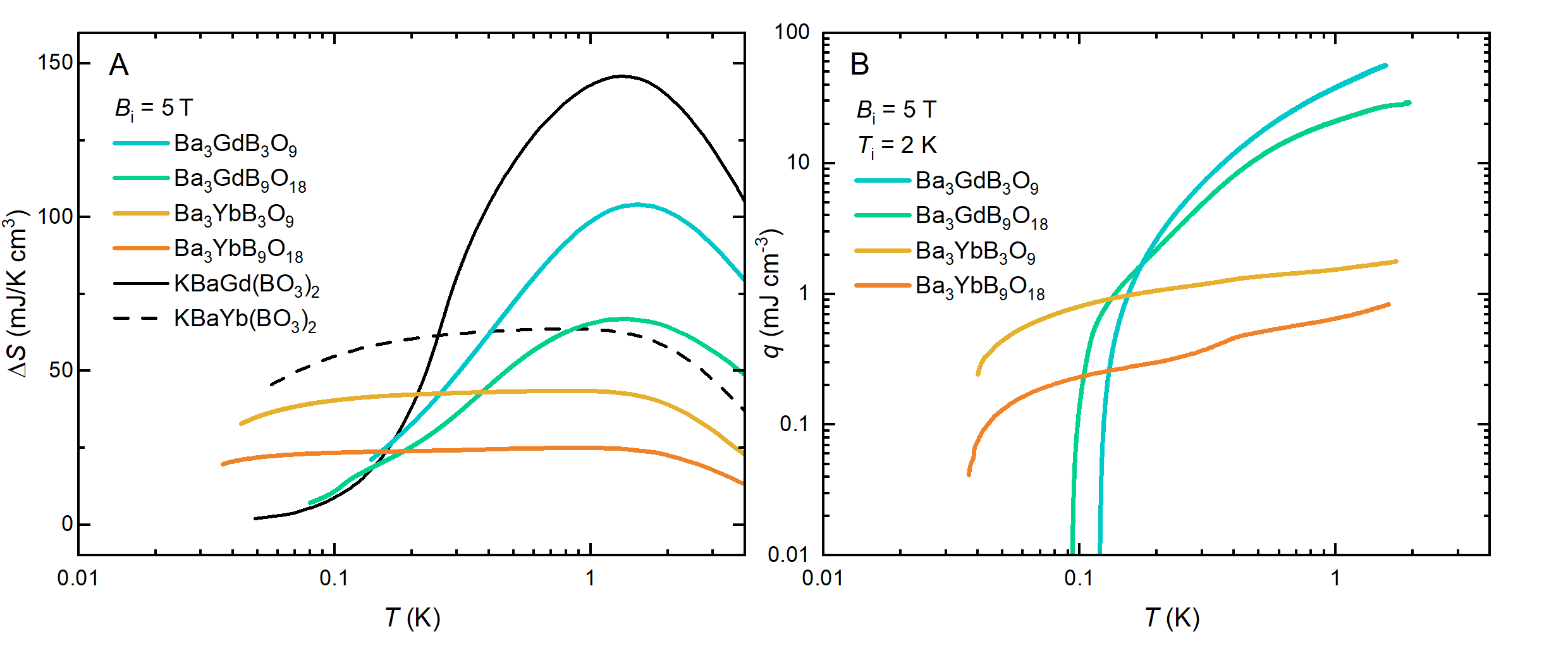}
\caption{Volumetric entropy increment $\Delta S(T) = S(T,0) - S(T,B_i)$ with $B_i = 5\,\mathrm{T}$ (A) as well as refrigerant capacity $q$ (panel B, see text) for different studied Yb- and Gd borates vs. temperature.  \label{fig:entropy_lift}}
\end{figure}
%The effective magnetic moment for both compounds is similar the effective magnetic moment density $\mu_\mathrm{eff}/V$ is significantly increased ($+\%\,\mu_\mathrm{eff}/V$) for \gt{}.
All substances reached significantly lower temperatures $T_\mathrm{ADR}$ during the ADR than their respective $|\Theta_\mathrm{W}|$ with all except \gn{} showing no signatures of long range order. This suggests a possible impact of geometrical frustration. However, more detailed study of the magnetic properties would be required to draw a clearer conclusion.

% Mathematische Werte:
% Yb3+ 4,54µB
% Gd3+ 7,94µB

To compare the refrigeration performance of various materials, the change in the volumetric magnetic entropy density $\Delta S = S_{B_i} - S_{B=0}$ was calculated for an initial field of $5\,\mathrm{T}$ (fig. \ref{fig:entropy_lift} A).
Likewise, the refrigerant capacity $q$ was calculated according to eq.~\ref{eq:ref-cap} for a demagnetization from $B_i = 5\,\mathrm{T}$ (fig. \ref{fig:entropy_lift} B).
This indicates the amount of heat the refrigerant can absorb during warm up to a specified temperature - in this case $2\,\mathrm{K}$.

In terms of entropy change (fig. \ref{fig:entropy_lift} A) the Yb (blue, green) and Gd (orange, yellow) compounds differ in magnitude as well as temperature distribution.
As expected, the Gd based substances offer significantly higher entropy lift per unit volume with \yt{} outperforming \yn{} due to the higher moment density.
However $\Delta S(T)$ shows a peak for these substances - this can be interpreted as the optimum operating conditions.
Meanwhile the Yb compounds offer a far broader plateau in $\Delta S(T)$ albeit at a significantly reduced magnitude.
The tendency for Yb based refrigerants to offer optimal performance over a wider temperature range has been observed for multiple mK refrigerants \cite{treu:2025}.

In comparison to the previously studied borates \kbayb{} and \kbagd{} the performance of the new refrigerants is limited \cite{tokiwa:2021, jesche:2023, treu:2025}.
\kbayb{} (fig. \ref{fig:entropy_lift} A dashed line) offers higher $\Delta S$ than \yt{} and \yn{} over the entire temperature range. %, while allowing for significantly lower minimal temperatures of below $22\,\mathrm{mK}$ \cite{tokiwa:2021}.
\gt{} and \gn{} outperform \kbagd{} (fig. \ref{fig:entropy_lift} A solid line) for very low temperatures $(T < 0.14\,\mathrm{K}$ and $T < 0.17\,\mathrm{K})$ but offer only limited entropy lift at higher temperatures.
This can be explained by the higher magnetic moment density in the diborates when compared to the triborates studied here.
Overall, enhanced magnetic moment density combined with strong frustration seems to be preferable to more dilute frustrated systems for refrigeration purposes.

\section{Conclusion}

Magnetic measurements, specific heat and ADR results for \yt{}, \yn{}, \gt{} and \gn{} were presented.
All materials exhibit slightly negative $\Theta_\mathrm{W}$ indicating weak antiferromagnetic interactions, which are stronger in case of the Gd-materials due to the larger moment size.
Magnetization of the ytterbium samples adheres closely to the single ion model, while the gadolinium systems necessitated a mean field approach.
The susceptibility data reveals paramagnetic behavior to $0.3\,\mathrm{K}$, underpinned by specific heat data clearly exhibiting a Schottky type anomaly.

Major MCE was observed in all substances during a demagnetization from $2\,\mathrm{K}, 5\,\mathrm{T}$.
Both \gn{} and \gt{} achieve low ADR temperatures for gadolinium systems while \yn{} and \yt{} are competitive with other rare earth oxide quantum magnets in regards to temperature.
All compounds offer significantly lower entropy densities and refrigerant capacities than the bench mark borates $\mathrm{KBa}(\mathrm{Gd}, \mathrm{Yb})(\mathrm{BO}_3)_2$.
For \gn{} the formation of long range magnetic order was observed by means of ADR warm up analysis.
The specific heat exhibits a clear lambda-shaped phase transition anomaly in zero field, followed by a broad maximum at temperatures above. Data in $1\,\mathrm{T}$ are well described by a $S=7/2$ Schottky anomaly.
Long range order has not been detected in \gt{} even though this compound has a higher magnetic moment density.
The absence of magnetic order in \gt{} is likely caused by frustration due to the trigonal lattice.
It also has to be considered, that this material has two nonequivalent magnetic sites.

Both \yn{} and \yt{} attain low temperatures during demagnetization when compared to other frustrated Yb based quantum magnets.
While they are competitive in regards to temperature they lack in terms of refrigerant capacity and entropy density when compared to incumbents like \kbayb{}.

\bibliography{your_external_BibTeX_file}% Produces the bibliography via BibTeX.

\end{document}